\documentclass
[preprint, showpacs,preprintnumbers,amsmath,amssymb,nofootinbib]{revtex4}
\usepackage{amssymb}
\usepackage{natbib}
\usepackage{graphicx}
\usepackage{dcolumn}
\usepackage{bm}

\newcommand{\bea}{\begin{eqnarray}}
\newcommand{\eea}{\end{eqnarray}}

\begin{document}
\title{ Testing the cosmic distance duality relation with Type Ia supernova and transverse BAO measurements }
\author{Min Wang$^1$, Xiangyun Fu$^1$\footnote{corresponding author: xyfu@hnust.edu.cn}, Bing Xu$^2$\footnote{corresponding author: xub@ahstu.edu.cn},
Yang Huang$^1$,
Ying Yang$^1$, and Zhenyan Lu$^1$}
\address{$^1$Department of Physics, Key Laboratory of Intelligent Sensors and Advanced Sensor Materials, Hunan University of Science and Technology, Xiangtan, Hunan 411201, China\\
$^2$School of Electrical and Electronic Engineering, Anhui Science and Technology University, Bengbu, Anhui 233030, China
}

\begin{abstract}
In this work, we test the cosmic distance duality relation (CDDR) by comparing the angular diameter distance (ADD) derived from the transverse Baryon Acoustic Oscillations (BAO) data with the luminosity distance (LD) from the Pantheon type Ia supernova (SNIa) sample. The binning method and Gaussian process are employed to match ADD data with LD data at the same redshift. First, we use nonparametric and parametric methods to investigate the impact of the specific prior values of the absolute magnitude $M_{\rm B}$ from SNIa observations and the sound horizon scale $r_{\rm s}$ from transverse BAO measurements on the CDDR tests. The results obtained from the parametric and non-parametric methods indicate that specific prior values of $M_{\rm B}$ and $r_{\rm s}$ lead to significant biases on the CDDR test. Then, to avoid these biases,  we propose a method independent of $M_{\rm B}$ and $r_{\rm s}$  to test CDDR by considering the fiducial value of $\kappa\equiv10^{M_{\rm B} \over 5}r_{\rm s}$ as a nuisance parameter and then marginalizing its influence with a flat prior in the analysis. No violation of the CDDR is found, and the transverse BAO measurement can be used as a powerful tool to verify the validity of CDDR in the cosmological-model-independent method.

$\mathbf{Keywords:}$  Cosmic distance duality relation, Parametrization, Cosmological-model-independent method
\end{abstract}

\pacs{ 98.80.Es, 95.36.+x, 98.80.-k}

 \maketitle

\section{Introduction}
The cosmic distance duality relation (CDDR) is the most fundamental relationship in modern cosmology, which states that the luminosity distance (LD) $D_{\rm L}$ and the angular diameter distance (ADD) $D_{\rm A}$ follow the following relation,
\begin{equation}
{D_{\rm L}\over D_{\rm A}}{(1+z)^{-2}}=1\,.
\end{equation}
Here, $z$ is the redshift of the observed source. This relation is first proved  by  Etherington~\cite{Etherington1993}, and it is based on two
fundamental hypotheses: (a) the photon always travels
along null geodesics in Riemannian geometry; (b) the
number of photons is conserved~\cite{ellis1971,ellis2007}. As a fundamental and crucial relation independent of Einstein
field equations and the nature of matter, CDDR is widely used to observe the large-scale distribution of galaxies and the near uniformity of the Cosmic Microwave Background (CMB) radiation temperature~\cite{Aghanim2020}, determine the gas mass density distribution and temperature distribution of galaxy clusters~\cite{Cao20112,Cao2016}, test the geometric shape of galaxy clusters~\cite{Holanda2011}, and measure the curvature of the universe using strong gravitational lensing systems~\cite{Liu2020,Xia2017,Qi2019}. In astronomical observations, any violation of the CDDR  implies the existence of new physics or the presence of unexplained errors in the observations. Therefore, it is necessary to test the validity of CDDR through various astronomical observations.

One can in principle check the validity of the CDDR with astronomical
observations by comparing the LD with
ADD of some objects at the same redshift. In the early stage, due to the lack of observational data, the CDDR test was performed by comparing the observed values with the corresponding
theoretical values in the assumed cosmological model. De Bernardis {\it et al.}~\cite{DeBernardis2006} and Uzan {\it et al.}~\cite{Uzan2004} tested CDDR using galaxy cluster samples with the LD $D_{\rm L}$ from the  cosmological constant $\Lambda$ cold dark matter ($\Lambda$CDM) model~\cite{Reese2002,Bonamenteet2006}, and found no violation
from the CDDR. Using galaxy cluster data and the conventional $\Lambda$CDM model constrained from WMAP (7 years) results, Holanda {\it et al.} validated the CDDR with the elliptical and spherical $\beta$ models at $1\sigma$ and $3\sigma$ confidence level (CL)~\cite{Holanda2011}, respectively.
Then, Lazkoz {\it et al.} verified the effectiveness of CDDR at $2\sigma$ CL by comparing the dark energy model constraints from the CMB and baryon acoustic oscillation (BAO) with those obtained from the latest type Ia supernova (SNIa) standard candle data at that time~\cite{Lazkoz2008}. However, any hypothetical cosmological model will bring bias to the testing of CDDR.

With the development of astronomic observations, CDDR  can be tested directly by comparing the LD of Type Ia supernovae with the ADD of other astronomical observations in the cosmological-model-independent method. With the galaxy cluster samples~\cite{Bonamenteet2006,DeFilippis05}  and  SNIa data, Holanda {\it et al.}~\cite{holanda20103} and Li {\it et al.}~\cite{Li2011} performed the CDDR test with adoption of a criterion ($\Delta z=|z_{\rm ADD}-z_{\rm SNIa}|<0.005$) and chose the closest one. To avoid the larger statistical errors  caused by using only one SNIa data point from all those available that meet the selection criterion, Meng {\it et al.}~\cite{Meng2012} did not use the nearest point from  SNIa compilation,  they binned the available data (hereafter referred to as binning method) to obtain LD that matches the corresponding ADD sample. They concluded that the marked triaxial ellipsoidal model can better describe the geometrical structure of galaxy cluster than the spherical model if the CDDR is valid. Then, Wu {\it et al.} tested CDDR by comparing the Union 2.1 compilation  with the five ADD data points from the BAO measurements, and found that the BAO measurement is a very powerful tool to test the CDDR due to the precision of the BAO measurements~\cite{Wu2015}.
Still some other tests, involving the ADD of galaxy clusters, current CMB observations~\cite{Ellis2013},  Hubble parameter data $H(z)$ from cosmic chronometers, gas mass fraction (GMF) measurements in  galaxy clusters~\cite{Goncalves2012,Goncalves2015},  strong gravitational lensing (SGL)~\cite{Cao2012,Cao2015}, South Pole Telescope (SPT) Sunyaev–Zel'dovich (SZ) clusters and
X-ray  Multi-mirror Mission (XMM)-Newton measurements~\cite{Bocquet2015,Bora2021,Benson2013,Bulbul2019,Liu2021,McDonald2016,Stalder2013},  the x-ray surface brightness
observations of galaxy clusters jointly with type Ia supernovae and CMB temperature~\cite{Holanda201711,Luzzi2009}, and time delay lenses~\cite{Balmes}, are performed to investigate the validity of the CDDR by assuming a deformed  CDDR, such as ${D_{\rm L}}{(1+z)}^{-2}/{D_{\rm A}}=\eta{(z)}$, in different redshift range, and the results show that CDDR is consistent with astronomic observations at certain CLs~\cite{avtidisgous,Holanda2012a,Santos2015,
Stern2010,Holanda2012,Liao2011,Holanda20171,Liao2016,fuxiangyun,Fu2017,Liang2013,Rana,Ruan,Xu2022,Holanda2017,Zheng2020}.

SNIa observations and BAO measurements, as standard candles and standard rulers in astronomical observations, play important roles in the CDDR test. It is worth noting that LD obtained from SNIa observation depends on its peak absolute magnitude $M_{\rm B}$, which is considered to be a fixed value that does not vary with other factors. Recently, using the Hubble $H(z)$ measurements from the cosmic chronometers, Kumar {\it et al.}
probed the $M_{\rm B}$ variability of SNIa
along with its correlation with spatial curvature and CDDR~\cite{Kumar2022}, and they found no evolution of $M_{\rm B}$ within $2\sigma$ CL. There have been some other attempts to derive the value of $M_{\rm B}$ from the cosmological perspective~\cite{Camarena2020,Dinda2023}, and different values of $M_{\rm B}$ are obtained from  the SNIa data with other combinations of datasets, such as CMB
observations, the cosmic chronometers data for the Hubble parameter, and BAO
observations. It was also found there was a mismatch between the SNIa absolute magnitude calibrated by Cepheids at
$z\leq 0.01$ and $z>0.01$~\cite{Camarena2021,Camarena20201}. Recently, hints of possible weak
evolution of $M_{\rm B}$ were also
pointed out in Refs.~\cite{Kazantzidis2021,Kazantzidis2020}. In addition, the so-called fitting problem remains a challenge for BAO peak location as a standard ruler~\cite{Ellis1987}, although BAO measurements are employed to analyze various cosmological parameters. In particular, Roukema {\it et al.} recently detected the environmental dependence of BAO location~\cite{Roukema2015,Roukema2016}. Moreover, Ding {\it et al.} and Zheng {\it et al.} pointed out a noticeable systematic difference between Hubble $H(z)$ measurements based on BAO and those obtained with differential aging techniques~\cite{Ding2015,Zheng2016}. Different values of sound horizon scales $r_{\rm s}$ and the present Hubble parameter $H_0$ are obtained from various observational data, such as CMB observations~\cite{Ade2016,Bennett2013}, the Sloan Digital Sky Survey (SDSS) data release 11 galaxies~\cite{Carvalho2020}, and BAO measurements~\cite{Verde2017}. Therefore, any CDDR tests completed by the priors of $M_{\rm B}$ and $r_{\rm s}$ are not absolutely independent of cosmological model, as the LD and ADD derived from $M_{\rm B}$ of SNIa observation and $r_{\rm s}$ of BAO measurements are  dependent on the cosmological model to some extent. In addition, any priors of  $M_{\rm B}$ and $r_{\rm s}$ may lead to biases in the CDDR tests. Therefore, it is meaningful to investigate the impacts of  $M_{\rm B}$ and $r_{\rm s}$ on the CDDR tests, and explore new methods that are independent of $M_{\rm B}$ and $r_{\rm s}$  to test CDDR, which is the main motivation of this work.

In this work, we perform the CDDR test by comparing LD from Pantheon SNIa data with ADD from transverse BAO data. The function $\eta(z)=D_{\rm L}(z)(1 + z)^{-2}/D_{\rm A}(z)$ is used to verify possible deviation at any redshift. Using the Gaussian process, we obtain the smoothing LD of SNIa data and the smoothing ADD from the transverse BAO data with priors of the absolute magnitude $M_{\rm B}$ and the sound horizon scale $r_{\rm s}$. We first investigate the impacts of $M_{\rm B}$ and $r_{\rm s}$ on CDDR tests by comparing the smoothing LD function with the ADD function in the nonparametric method. Then, with three parametrization for the function $\eta(z)$, we test CDDR by using the binning method to match the SNIa data with the BAO measurements at the same redshift in the parametric method. The results obtained from the nonparametric and parametric methods show that the priors of $M_{\rm B}$ and $r_{\rm s}$ may cause significant biases in the  CDDR test. To avoid these bias, we consider variables $M_{\rm B}$ and $r_{\rm s}$ as nuisance parameters to determine the LD and ADD, and then adopt a new variable $\kappa\equiv10^{M_{\rm B} \over 5}r_{\rm s}$ to analytically marginalize its influence with a flat prior in the analysis. We show that CDDR is consistent with the observed data, and the parametric method for testing CDDR is independent of cosmological models.

\section{Data and Methodology}

\subsection{Data}
To validate CDDR, two types of cosmological distances are usually required, i.e., LD $D_{\rm L}$ and ADD
$D_{\rm A}$. In this work, LD data are obtained from the Pantheon SNIa observations,
which consists of 1048 data points from the Pan-STARRS1 Deep Survey with the redshift range of $0.01<z<2.26$~\cite{Scolnic2018}. The distance modulus of the Pantheon compilation is calibrated from the SALT2 light-curve fitter through applying the Bayesian Estimation Applied to
Multiple Species with Bias Corrections method to determine
the nuisance parameters and taking account of the distance
bias corrections, namely, $\mu = m_{\rm B}-M_{\rm B}$. Here $m_{\rm B}$ denotes the peak apparent magnitude observed in the rest-frame B-band, and $M_{\rm B}$ represents the absolute magnitude. The observational constraints on
the cosmological parameters, such as the deceleration parameter, the matter-energy density parameter, and the dark energy density parameter, etc. are estimated based on the
value of ${m_{\rm B}}$ and $M_{\rm B}$~\cite{Linden2009,Camarena2020}. SNIa observations firstly revealed the late time cosmic acceleration~\cite{Riess1998,Perlmutter1999,Scolnic2018}, where
the observations are based on the fact that SNIa can be considered as a standard candle and its peak absolute magnitude, $M_{\rm B}$, is uniform. However the constancy of absolute magnitude $M_{\rm B}$ of SNIa has often been questioned as an immediate
retort to the $H_0$ tension.

In recent years, several studies have raised the question of whether or not $M_{\rm B}$ might
evolve with redshift. With an inverse distance ladder, the CMB constraint on the sound horizon
predicts $M_{\rm B}\sim -19.4 {\rm mag}$, while the estimate from SH0ES corresponds to $M_{\rm B}\sim-19.2{\rm mag}$.  Hence, we first focus on whether the difference in $M_{\rm B}$ would affect the CDDR test. In this work, we consider two specific priors of $M_{\rm B}$ derived from different observational data sets within various redshift ranges, namely, (a) $M_{\rm B}^{\rm D20}={-19.23\pm0.0404 {\rm mag}}$, which was
derived from Supercal supernovae in the relative low redshift range $0.023< z < 0.15$ by Camarena and Marra~\cite{Camarena2020} in $\Lambda$CDM through a de-marginalization
of the SH0ES determination~\cite{Reid2019} (hereinafter referred to as $M_{\rm B}^{\rm D20}$); (b) $M_{\rm B}^{\rm B23}={-19.396\pm0.016{\rm mag}}$ obtained by combining SNIa observations with BAO observations~\cite{Dinda2023} (hereinafter referred to as $M_{\rm B}^{\rm B23}$). Considering the observational uncertainty of $M_{\rm B}$, the error bar of $\mu$  can be written as $\sigma_\mu=\sqrt{\sigma_{m_{\rm B}}^2+\sigma_{M_{\rm B}}^2}$.
The relation between LD $D_{\rm L}$~\cite{Zhou2019} and distance modulus $\mu$ can be expressed as
\begin{equation}
\mu(z)=5\log_{10}[D_{\rm L}(z)]+25\,,
\end{equation}
and the uncertainty of $D_{\rm L}$ can be obtained from the following formula $\sigma_{D_{\rm L}}={\ln{10}}D_{\rm L}\sigma_{\mu}$/5.

The measurement of the BAO scale is usually obtained by applying the two-point correlation function (2PCF) to a large number of galaxy distributions, where the BAO feature appears as a bump at the corresponding scale. This type of analysis may bias cosmological parameter constraints, since it assumes a fiducial cosmology in order to convert measured angular positions and redshifts into comoving distances. The astronomic observation of the transverse BAO scale can be obtained by using the 2-point angular correlation function (2PACF), which involves only the angular separation $\theta$ between pairs~\cite{Carvalho2020}. This yields the ADD information almost model-independently, provided that the comoving sound horizon is known~\cite{Carvalho2020,Nunes2020}. The 15 transverse BAO data points in this work are taken from Table I in Ref.~\cite{Nunes2020}, which were released  publicly from the SDSS~\cite{York2000}.
The transverse BAO data set was also employed to test the cosmological principle~\cite{Arjona2021,Bengaly2022,Wang2023}.
The ADD $D_{\rm A}$ can be obtained from the transverse BAO data,
\begin{equation}
D_{\rm A}(z)={r_{\rm s}\over{(1+z)}\theta_{\rm BAO}}\,.
\end{equation}
Here, $r_{\rm s}$ is the sound horizon scale at the drag epoch, and $\theta_{\rm BAO}$ corresponds to the comoving observed angle from the transverse BAO measurements.

However, the value of sound horizon scale $r_{\rm s}$ is dependent on the Hubble constant (as a local anchor of the
cosmic distance scale), and different values of $r_{\rm s}$ are obtained from various astronomic observations. Recently,  the sound horizon scale $r_{\rm s}$  was calibrated  from CMB observations in the high redshift region $z>1000$, i.e., $r_{\rm s}={147.33\pm0.49{\rm Mpc}}$ and $r_{\rm s}={152.30\pm1.30{\rm Mpc}}$ from the
most recent Planck~\cite{Ade2016} and WMAP9~\cite{Bennett2013} measurements,
respectively. In addition, using the SDSS data release 11
galaxies and the prior of the matter density parameter given
by the SNIa data, Carvalho {\it et al.}
obtained constraint on $r_{\rm s}$ (hereafter referred to as $r_{\rm s}^{{\rm C}20}$) at the relatively low redshift~\cite{Carvalho2020},
\begin{equation}
\label{rs1}
r_{\rm s}^{\rm C20}={107.4\pm1.7\,h^{-1}{\rm Mpc}}.
\end{equation}
Here,  $h$ is the Hubble constant $H_0$ in units of $100 {\rm km\,s^{-1}Mpc^{-1}}$.
Verde {\it et al}. also obtained the
sound horizon $r_{\rm s}$ (referred to as $r_{\rm s}^{\rm V17}$) with SNIa  and BAO measurements~\cite{Verde2017},
\begin{equation}
\label{rs2}
r_{\rm s}^{\rm V17}={101.0\pm2.3{\, h^{-1}}{\rm Mpc}}\,.
\end{equation}
We adopt the prior values of the
sound horizon $r_{\rm s}^{\rm C20}$ and $r_{\rm s}^{\rm V17}$, which are obtained from different astronomic observations at the relatively low redshifts, to further investigate the impact of $r_{\rm s}$ on the CDDR test. In fact, the value of $r_{\rm s}^{\rm V17}$ is also related to the observed value of the Hubble constant $H_0$. However, there are differences in the measurements of the Hubble constant from astronomic observations, known as the Hubble tension (the discrepancy between the local measurements based on cepheid variables and SNIa and CMB analysis within the context of the $\Lambda$CDM model). For simplicity and the consistency of observational SNIa data, we use the Hubble constant $H_0={73.04\pm1.04{\rm km\,s^{-1}\,Mpc^{-1}}}$, which was obtained from a Cepheid-only calibration
of $42$ SNIa data points with good quality~\cite{Riess2022}, to investigate the impacts of $M_{\rm B}$ and $r_{\rm s}$ on the CDDR tests.
\subsection{Gaussian process and  Binning method}
To test the validity of CDDR, the direct method is to make the comparison between ADD and LD from various observations at the same redshift. Due to the lack of the astronomic observational data, different methods are employed to match ADD with LD at the same redshift. Here, we employ two cosmological model-independent methods, namely the Gaussian process and  the binning method, to perform this task.

\begin{figure}[htbp]
\includegraphics[width=6cm]{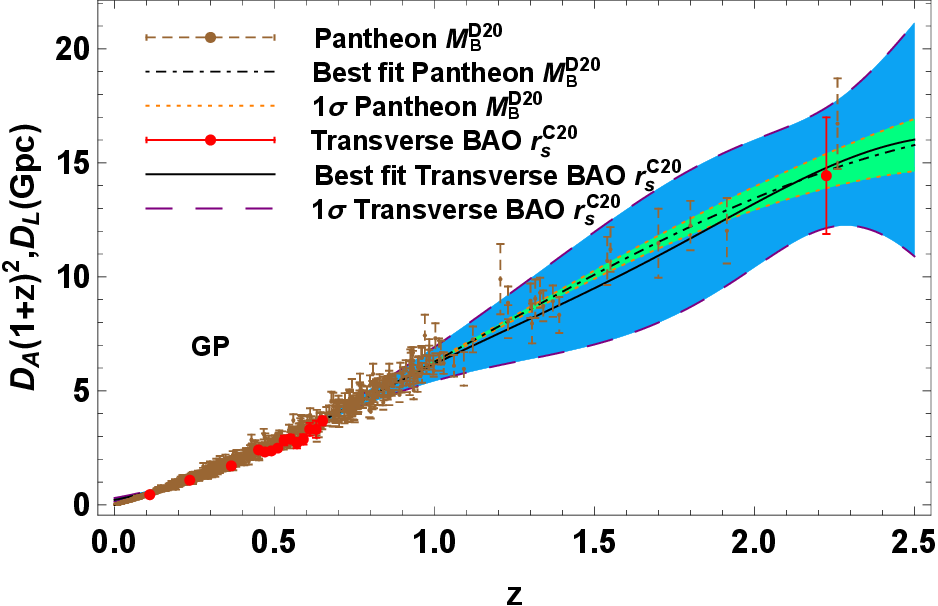}
\includegraphics[width=6cm]{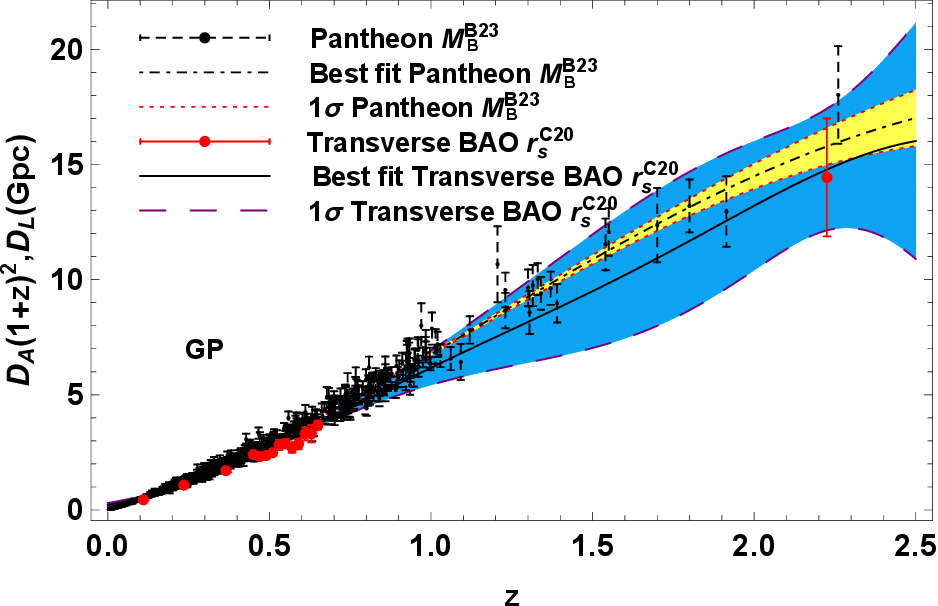}
\includegraphics[width=6cm]{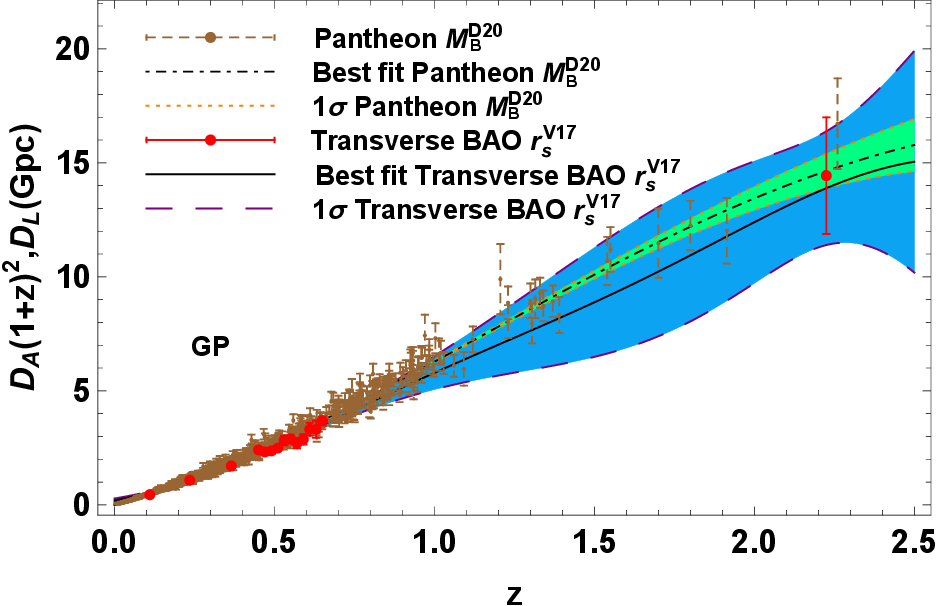}
\includegraphics[width=6cm]{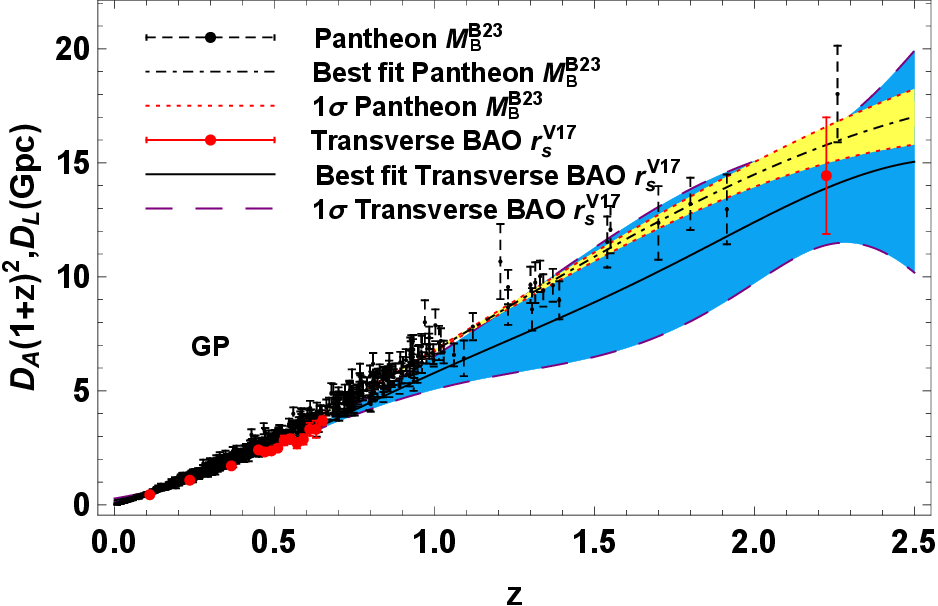}
\caption{\label{figdl1}  The distribution of LD $D_{\rm L}$ from the observed  Pantheon and $D_{\rm A}(1+z)^2$ from the transverse BAO, and the corresponding reconstructed curves with the priors of $M_{\rm B}^{\rm D20}$ (left panel), $M_{\rm B}^{\rm B23}$  (right panel), $r_{\rm s}^{\rm C20}$ (upper panel), and $r_{\rm s}^{\rm V17}$ (bottom panel), respectively. }
\end{figure}

We reconstruct the LD or ADD as a smoothing function from SNIa observations or transverse BAO measurements with the Gaussian process~\cite{Seikel2012,Carlos2020,Shafieloo2012}.
Gaussian process can be used for reconstructing a function $f(x)$ given function values $x$ under the assumption of probability distribution, assuming a probability distribution for finite-dimensional random variables. It also employs a kernel function and covariance function $k(x,x')$.  Different choices for kernels may lead to  slightly more degenerate reconstruction, thereby reducing
the statistical significance of these deviations~\cite{Bengaly2022}. More recently, Bayes factors were used to evaluate
the differences between different kernels with the cosmic chronometer data, SNIa, and gamma
ray burst, and it showed that Bayes factors indicate no
significant dependence of the data on the kernels~\cite{Zhang2023}. Here, we ignore the influences of different kernels and choose the general covariance kernel, namely, the squared exponential~\cite{Seikel2012},
\begin{equation}
k(x,x')={\sigma^{2}_f{\rm exp}{(-{{(x-x')^2}\over{2l^{2}}})}}\,,
\end{equation}
$\sigma_f$ and $l$ are the hyper-parameters in the fit. $l$ represents a measure of the correlation length of the correlation that provides the width of the reconstructed curve, $\sigma_f$ denotes the typical variation in function values, and $x$ and $x'$ represent two different points in space~\cite{Seikel2012}.
GP has been widely applied in cosmology, such as for tests on cosmological constants with supernovae~\cite{Yahya2014}, constraints on the dark energy equation of state~\cite{Seikel2012} and cosmological mixing parameters~\cite{Mukherjee2021,Holanda2013}, tests on the Cosmological Principle~\cite{Wang2023}, and inference of the Hubble constant~\cite{Li2016,Verde2014,Busti2014}.
The reconstructed results derived from different priors of $M_{\rm B}$ are shown in Fig.~\ref{figdl1}.

To verify the validity of CDDR, instead of choosing the nearest SNIa data
to match SNIa observation with ADD measurements at the same redshift with a selection criterion
$\Delta z=|z_{\rm ADD}-z_{\rm SNIa}|<0.005$ proposed in Refs.~\cite{holanda20103,Li2011,Liao2016}, we bin the available data from SNIa data points that meet the selection criterion. This method, named the binning method, may be used to  avoid the statistical errors resulting from merely one SNIa data point with all those available that
meet the selection criterion, and has been used to discuss
the CDDR tests in Refs.~\cite{Wu2015,Meng2012}. Here, we adopt an inverse variance weighted average of all the selected data, and select the LD data points from SNIa observation  through a procedure. Once the data points have been matched with the BAO sample, they will no longer be used to avoid  the SNIa-to-SNIa correlations between individual CDDR test.
The weighted mean LD $\bar{D_{\rm L}}$ and its uncertainty $\sigma_{\bar{D_{\rm L}}}$ can be obtained  with conventional
 data reduction techniques in Chapter.($4$) of the Ref.~\cite{Bevington},
\begin{equation}
\label{avdi1}
\bar{D_{\rm L}}={\sum(D_{{\rm L}i}/\sigma_{D_{{\rm L}i}}^2)\over \sum1/\sigma_{D_{{\rm L}i}}^2},
\end{equation}
\begin{equation}
\label{erroravdi1}
\sigma^2_{\bar{D_{\rm L}}}={1\over \sum1/\sigma_{D_{{\rm L}i}}^2}\,.
\end{equation}
Here, $D_{{\rm L}i}$ denotes the $i$th appropriate luminosity distance data points, and $\sigma_{D_{{\rm L}i}}$ corresponds its observational uncertainty.

\begin{figure}[htbp]
\includegraphics[width=6cm]{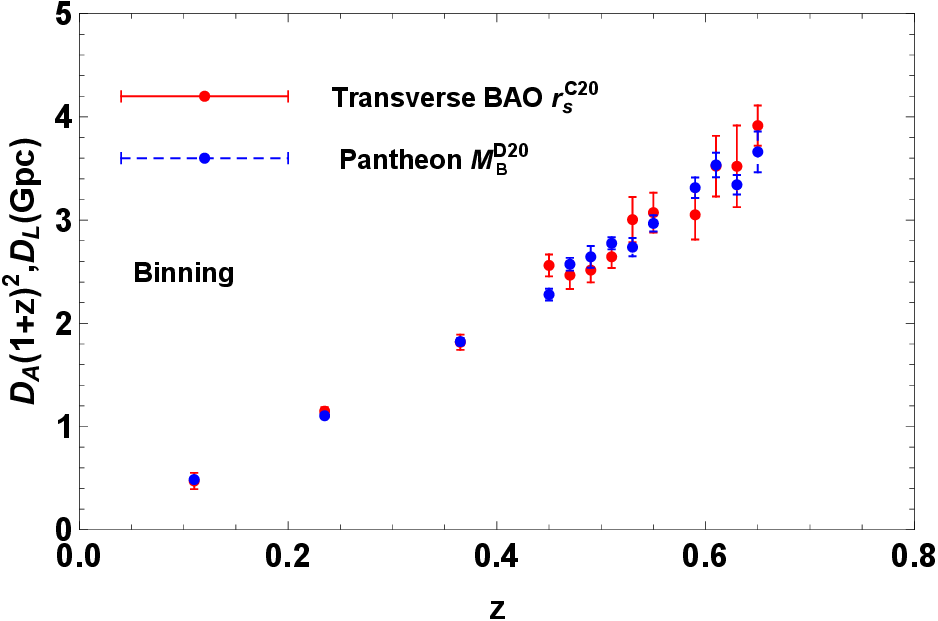}
\includegraphics[width=6cm]{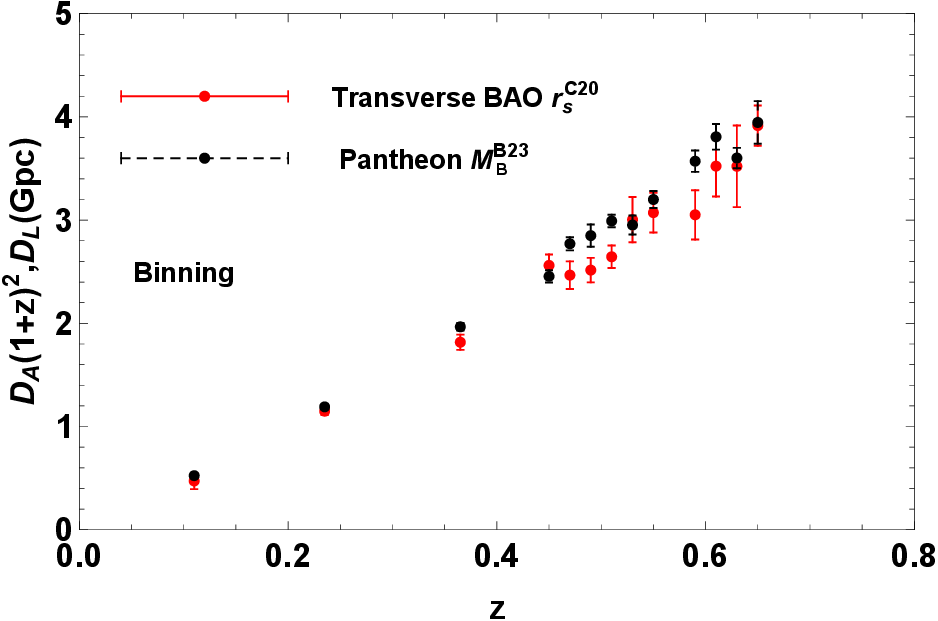}
\includegraphics[width=6cm]{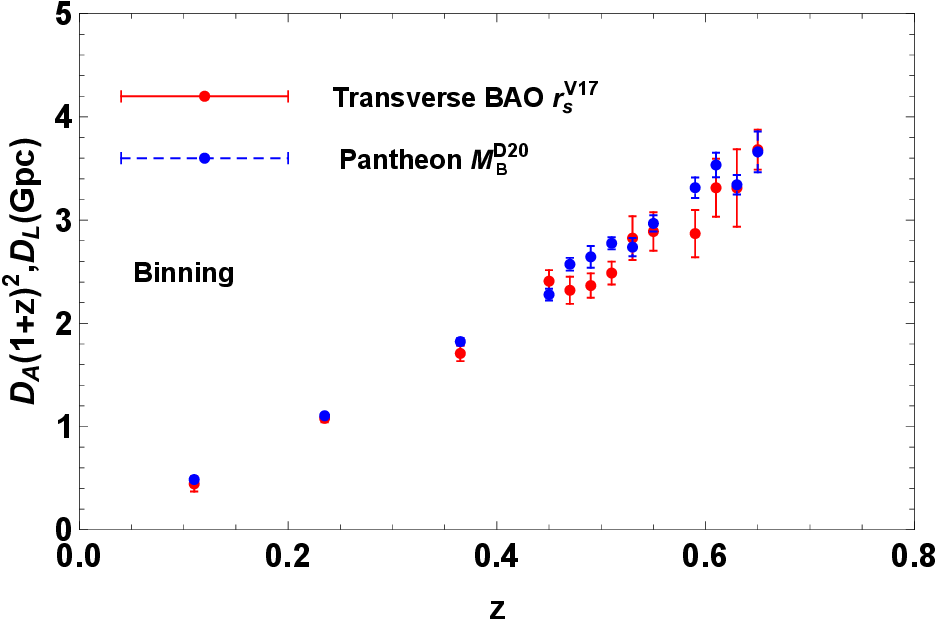}
\includegraphics[width=6cm]{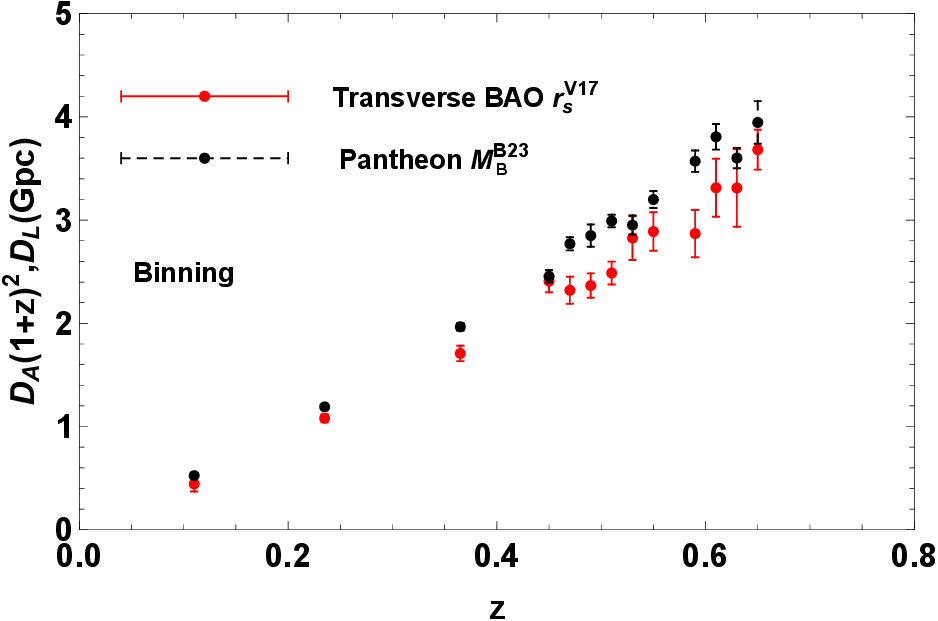}
\caption{\label{figdl2}  The sample catalogs of the observed  $D_{\rm A}(1+z)^2$ distribution from the 13 transverse BAO data points and the corresponding LD $D_{\rm L}$ from Pantheon data obtained from the binning method with the priors of $M_{\rm B}^{\rm D20}$ (left panel), $M_{\rm B}^{\rm B23}$ (right panel), $r_{\rm s}^{\rm C20}$ (upper panel), and $r_{\rm s}^{\rm V17}$ (bottom panel), respectively. }
\end{figure}

In this work, we choose Pantheon SNIa data points and match them with the transverse BAO data points, as the number of Pantheon SNIa far exceeds the number of BAO measurements. Only 13 data points in BAO measurements meet the selection criteria. Data points at $z=0.57$ and $z=2.225$ are discarded. The distribution of the transverse BAO data and SNIa data is shown in Fig.~\ref{figdl2}.

\subsection{Methodology}
\begin{figure}[htbp]
\includegraphics[width=6cm]{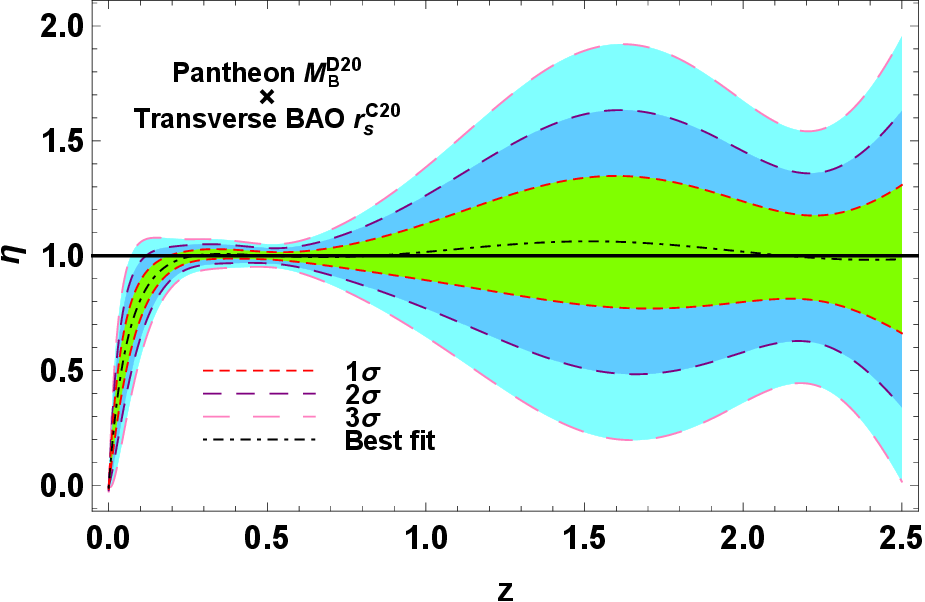}
\includegraphics[width=6cm]{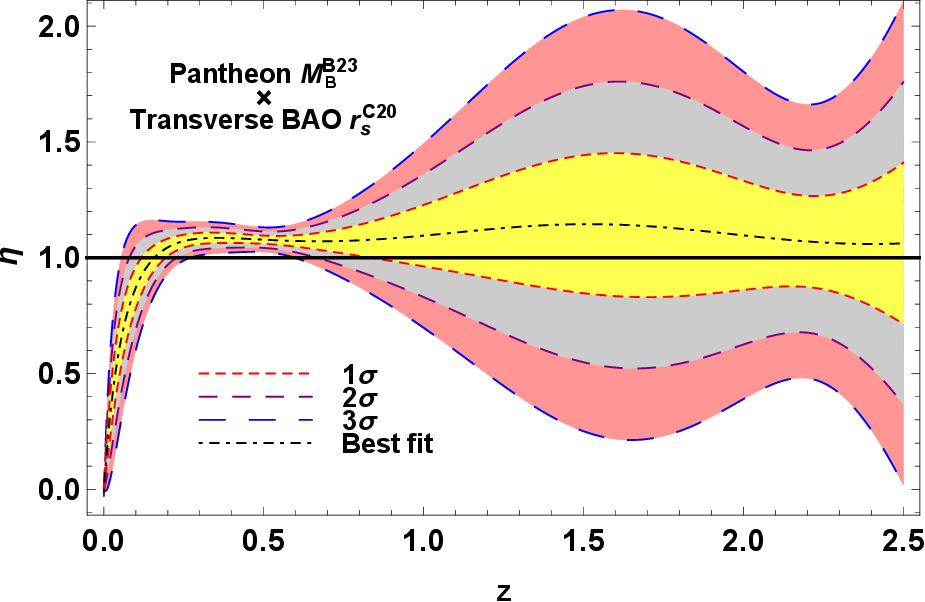}
\includegraphics[width=6cm]{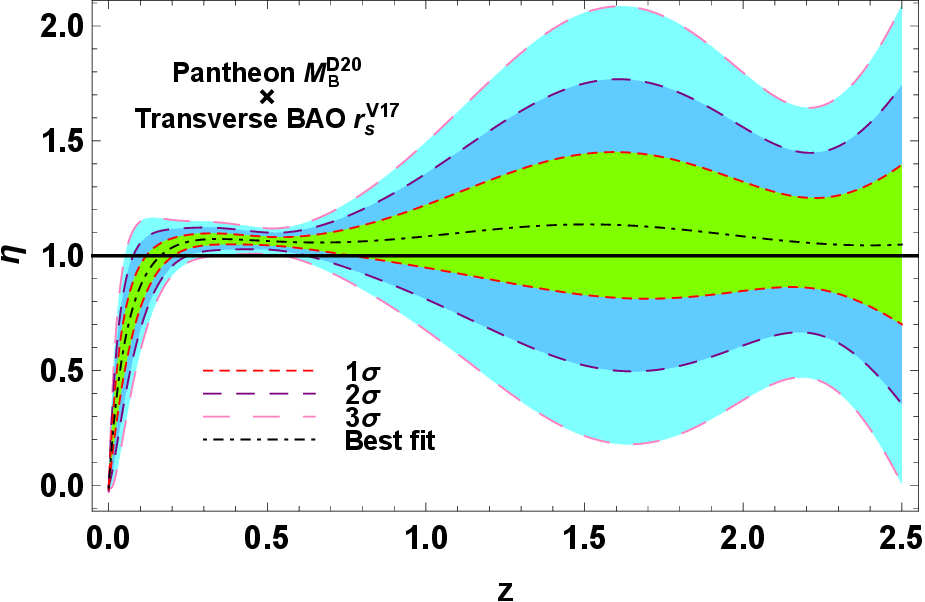}
\includegraphics[width=6cm]{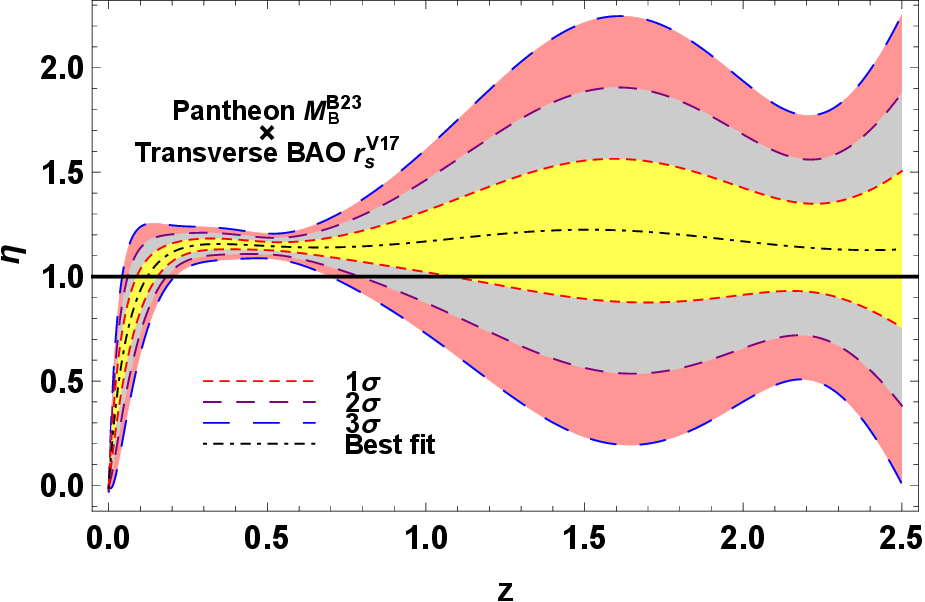}
\caption{\label{figeta1}  The distribution of $\eta(z)$ obtained from Gaussian process with the priors of $M_{\rm B}^{\rm D20}$ (left panel), $M_{\rm B}^{\rm B23}$  (right panel), $r_{\rm s}^{\rm C20}$ (upper panel) and $r_{\rm s}^{\rm V17}$ (bottom panel) with the nonparametric method. }
\end{figure}

We adopt the $\eta(z)$ function to verify the possible deviations from CDDR at any redshifts by comparing the $D_{\rm L}$ from  SNIa with the $D_{\rm A}$ from transverse BAO measurements. $\eta(z)$ can be obtained with the following expression,
\begin{equation}\label{eta}
\eta(z)={D_{\rm L}\over D_{\rm A}}{(1+z)^{-2}}\,.
\end{equation}
At any redshift, $\eta(z)\neq 1$  implies a discrepancy between CDDR and the astronomic observations.
We first obtain the function $\eta(z)$  at any redshifts by comparing the smoothing ADD $D_{\rm A}(z)$ with the LD $D_{\rm L}(z)$ at the same redshift, where $D_{\rm A}(z)$ and $D_{\rm L}(z)$ are obtained with different priors  $M_{\rm B}$ and $r_{\rm s}$.   It is obvious that this method of the CDDR test is nonparametric, and the results are shown in Fig.~\ref{figeta1}.

The parametric method plays an important role in testing CDDR. We adopt three parameterizations for $\eta(z)$, namely a linear form P1: $\eta(z)=1+\eta_0z$ and two non-linear forms P2: $\eta(z)=1+\eta_0z/(1+z)$, and P3: $\eta(z)=1+\eta_0\ln(1+z)$.
Any $\eta_0\neq 0$ denotes the deviation between CDDR and the observational data. The observed $\eta_{\rm obs} (z)$ is obtained from Equ.~\ref{eta}, and the corresponding error can be written as
\begin{equation}\label{SGL}
\sigma^2_{\eta_{\rm obs}}=\eta^2_{\rm obs}\left[\left({\sigma_{D_{\rm A}(z)}\over{D_{\rm A}(z)}}\right)^2+\left(\sigma_{D_{\rm L}(z)}
\over{D_{\rm L}(z)}\right)^2\right]\,.
\end{equation}

\begin{figure}[htbp]
\includegraphics[width=6cm]{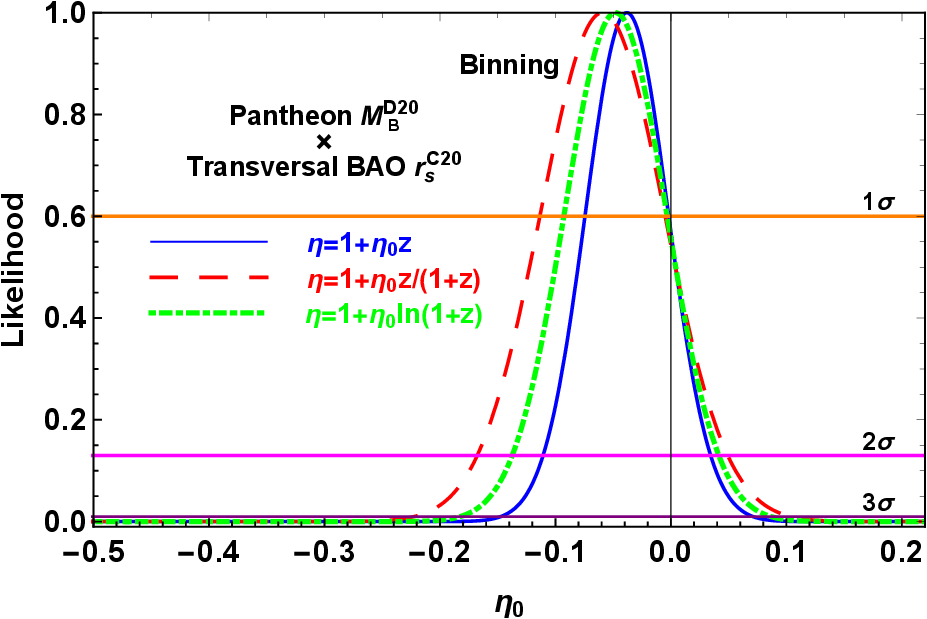}
\includegraphics[width=6cm]{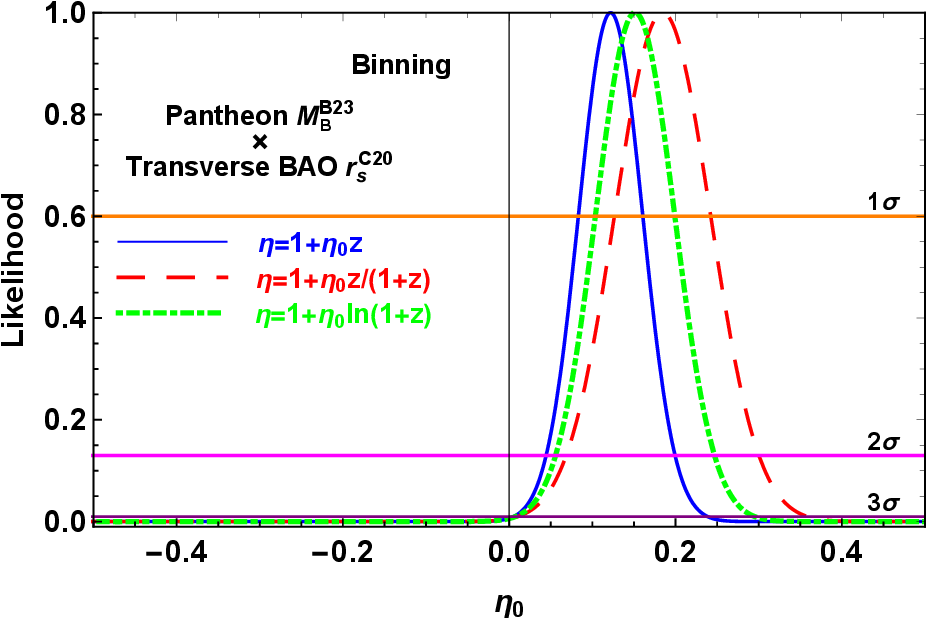}
\includegraphics[width=6cm]{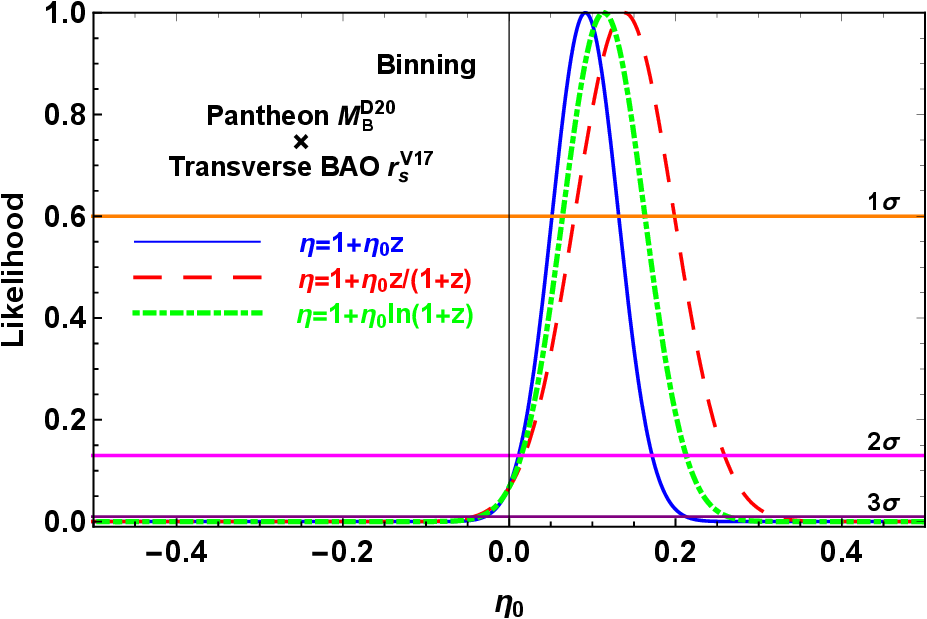}
\includegraphics[width=6cm]{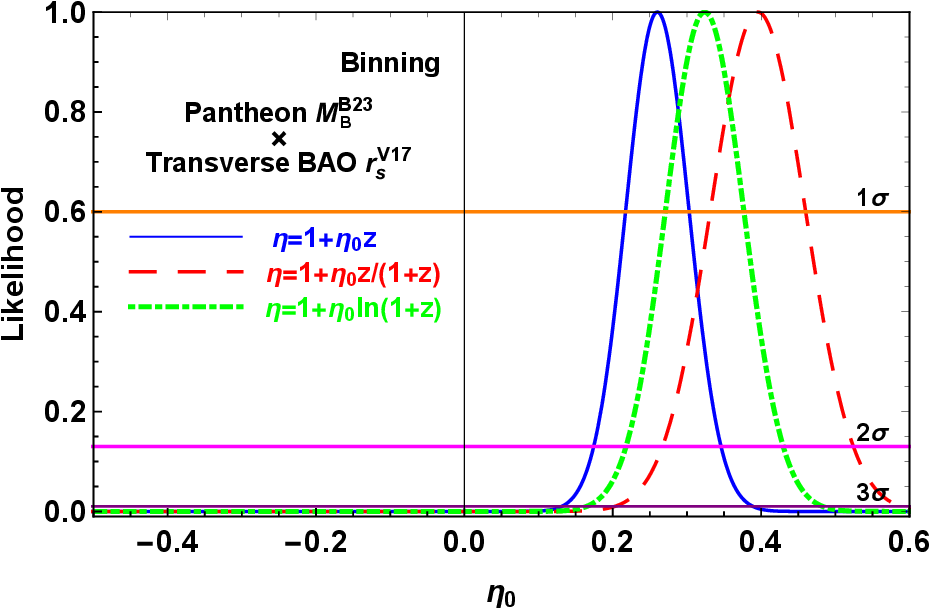}
\caption{\label{figeta} The likelihood distribution functions obtained with the priors of $M_{\rm B}^{\rm D20}$ (left panel), $M_{\rm B}^{\rm B23}$  (right panel), $r_{\rm s}^{\rm C20}$ (upper panel) and $r_{\rm s}^{\rm V17}$ (bottom panel) in parametric method. }
\end{figure}

Thus,  we obtain
\begin{equation}
\label{chi}
\chi^{2}(\eta_0)=\sum_i^{N}\frac{{\left[\eta(z)-\eta_{{\rm obs},\,i}(z) \right] }^{2}}{\sigma^2_{\eta_{{\rm obs},i}}}\,.
\end{equation}
Here, $N$ denotes the number of the available transverse BAO data points with the binning method.
The results of constraints on $\eta_0$ are shown in Fig.~\ref{figeta} and Tab.~\ref{likelihood1}.
\begin{table}[htp]
\begin{tabular}{ccccc}
\hline\hline
\scriptsize{parmetrization }  & \   P1: $1+\eta_0 {z}$\ \ &P2: $1+\eta_0 {z\over(1+z)}$  & \   P3: $1+\eta_0 {\ln(1+z)}$ \\
\scriptsize{$\eta_0^{\rm {A\,C}}\tiny$} & \scriptsize{${-0.038\pm0.037\pm0.072\pm0.109}$}& \scriptsize{${-0.059\pm0.055\pm0.109\pm0.165}$} & \scriptsize{${-0.048\pm0.046\pm0.090\pm0.137}$} \\
\scriptsize{$\eta_0^{\rm {B\,C}}\tiny$}  & \scriptsize{${0.092\pm0.039\pm0.081\pm0.119}$}& \scriptsize{${0.139\pm0.060\pm0.122\pm0.183}$} & \scriptsize{${0.114\pm0.050\pm0.099\pm0.148}$} \\
\scriptsize{$\eta_0^{\rm {A\,D}}\tiny$} & \scriptsize{${0.122\pm0.039\pm0.079\pm0.118}$}& \scriptsize{${0.184\pm0.059\pm0.117\pm0.176}$} & \scriptsize{${0.151\pm0.047\pm0.096\pm0.150}$} \\
\scriptsize{$\eta_0^{\rm {B\,D}}\tiny$} & \scriptsize{${0.260\pm0.044\pm0.087\pm0.129}$}& \scriptsize{${0.396\pm0.065\pm0.129\pm0.190}$} & \scriptsize{${0.324\pm0.053\pm0.106\pm0.162}$} \\
\scriptsize{$\eta_0\tiny\star$} & \scriptsize{${0.041\pm^{0.123}_{0.109}\pm^{0.256}_{0.211}\pm^{0.412}_{0.306}}$}& \scriptsize{${0.082\pm^{0.246}_{0.214}\pm^{0.525}_{0.404}\pm^{0.836}_{0.558}}$} & \scriptsize{${0.059\pm^{0.174}_{0.159}\pm^{0.370}_{0.291}\pm^{0.584}_{0.422}}$} \\
\hline\hline
\end{tabular}
\caption{Maximum likelihood estimation results for the parameterizations.  The $\eta_0$ is represented by the best fit value $\eta_{0,\rm{best}}\pm 1\sigma\pm 2\sigma\pm 3\sigma$ for each dataset. $1\sigma$, $2\sigma$, and $3\sigma$ denote the $68.3\%$, $95.4\%$, and $99.7\%$ CL, respectively.  The superscripts A, B, C, and D represent the cases obtained from $M_{\rm B}^{\rm D20}$, $M_{\rm B}^{\rm B23}$, $r_{\rm s}^{\rm C20}$, and $r_{\rm s}^{\rm V17}$, respectively. The superscript $\star$ denotes the results obtained from the flat prior for $\kappa$.}
\label{likelihood1}
\end{table}

It is easy to see that the results obtained from the nonparametric and parametric methods depend on the priors of the  $M_{\rm B}$ and $r_{\rm s}$. To test CDDR with a method independent of absolute magnitude $M_{\rm B}$ and sound horizon scale $r_{\rm s}$, we consider fiducial values of $M_{\rm B}$ and $r_{\rm s}$ as nuisance parameters
to determine LD $D_{\rm L}$ and  ADD $D_{\rm A}$, and marginalize their influences with
flat priors in the analysis. The likelihood distribution $\chi^{\prime 2}$ can be rewritten as
\begin{equation}
\label{chi2}
\chi^{\prime\,2}(\eta_0, \kappa)= \sum_i^{N}\frac{{{\alpha_i^2 \over \beta_i^2}{\kappa}^2- 2 {\alpha_i \over m_i}{\kappa}+1  }}{\sigma^{\prime\,2}_{{\eta_{{\rm obs},i}}}}\,.
\end{equation}
Here, $\alpha_i=\eta(z_i)$, $\beta_i=10^{({m_{{\rm B},i}\over 5}-5)}\theta_{{\rm BAO},i}(1+z_i)^{-1}$, $\kappa=10^{M_{\rm B} \over 5}r_{\rm s}$, and
\begin{equation}
\label{sigma01}
\sigma_{\eta_{{\rm obs},i}}^{\prime\,2}=\left({\ln{10}\over {5}}{\sigma_{m_{{\rm B},i}}}\right)^2+\left({\sigma_{\theta_{{\rm BAO},i}}\over{\theta_{{\rm BAO},i}}}\right)^2\,.
\end{equation}

Because the uncertainties of individual SNIa or the BAO measurements do not depend on $M_{\rm B}$ or $r_{\rm s}$, it is possible to remove these parameters from our fits
by analytically marginalizing over them. Therefore, following the method described in Ref.~\cite{Xu2022,Wang2023,Conley2011}, we marginalize analytically the likelihood
function over $\kappa$, and obtain
\begin{equation}
\label{chi3}
\chi_{\rm M}^{\prime\,2}(\eta_0)= C-{B^2\over {A}}+\ln{A\over 2\pi}\,,
\end{equation}
where $A=\sum \alpha_i^2/(\beta_i^2{\sigma^{\prime\,2}_{{\eta_{{\rm obs},i}}}})$, $B=\sum \alpha_i/(\beta_i{\sigma^{\prime\,2}_{{\eta_{{\rm obs},i}}}})$, and $C=\sum 1/{\sigma^{\prime\,2}_{{\eta_{{\rm obs},i}}}}$.
It is easy to see that the $\chi_{\rm M}^{\prime\,2}$ in Equ.~\ref{chi3} is  independent of the parameters $M_{\rm B}$ and $r_{\rm s}$. The results are shown in Fig.~\ref{figeta2} and Table~\ref{likelihood1}. It is worth noting that, the CDDR test in this work is also independent of the Hubble constant $H_0$, as $r_{\rm s}$ is marginalized in the analysis. To assess the ability of transverse BAO measurements to constrain parameters $\eta_0$,  we also include the results obtained from other observational data in Table~\ref{otherobser}.

\begin{table*}[htb]
\label{tab:results}
\begin{center}
\begin{tabular}{ccccc} \hline \hline
 Dataset used & P1: $1+\eta_0 {z}$\ \ &P2: $1+\eta_0 {z\over(1+z)}$  & \   P3: $1+\eta_0 {\ln(1+z)}$ \\ \hline
  ${\rm {Union2 }}+{\rm {Galaxy}}{\rm (Prior)}$~\cite{Li2011} & ${-0.22{\pm{0.11}}}$ & ${-0.33{\pm{0.16}}}$    \\ [1ex]
 ${\rm {Union2.1 }}+{91{\rm{GMF}}}{\rm (Prior)}$~\cite{Goncalves2015} & ${-0.08{\pm^{0.11}_{0.10}}}$ &     \\ [1ex]
 ${\rm {Union2.1}}+{\rm {BAO}} {\rm (Prior)}$ ~\cite{Wu2015}& ${-0.086{\pm{0.064}}}$ &  ${-0.131{\pm{0.098}}}$    \\ [1ex]
 ${\rm {Union2.1}}+{\rm {BAO}} {\rm (Marg)}$~\cite{Wu2015} & ${-0.174{\pm^{0.253}_{0.199}}}$ & ${-0.409{\pm^{0.529}_{0.381}}}$    \\ [1ex]
 ${\rm {Y_{SZ}}}-{\rm {Y_{X}}}\, {\rm {ratio}}+H(z){\rm (Prior)}$~\cite{Bora2021} & ${0.008{\pm{0.05}}}$ & ${0.019{\pm{0.11}}}$ & ${0.013{\pm{0.07}}}$   \\ [1ex]
 ${\rm {GMF}}+{\rm {SNIa}}+T_{\rm CMB}{\rm (Prior)}$~\cite{Holanda2017} & ${-0.020{\pm{0.027}}}$ &  ${-0.041{\pm{0.042}}}$    \\ [1ex]
 ${\rm {SNIa }}+{\rm {BAO}}{\rm (Marg)}$~\cite{Xu2020} & ${-0.07{\pm{0.12}}}$ & ${-0.20{\pm{0.27}}}$  & ${-0.12{\pm{0.18}}}$  \\ [1ex]
 ${\rm {SNIa }}+{\rm {BAO}}{\rm (Marg)}$~\cite{Xu2022} & ${-0.037{\pm^{0.110}_{0.097}}}$ & ${-0.101{\pm^{0.269}_{0.225}}}$  & ${-0.061{\pm^{0.173}_{0.149}}}$  \\ [1ex]
\hline \hline
\end{tabular}
\caption[]{ Summary of the constraints on parameter $\eta_0$ with different data sets. ``Prior"  represents the results obtained using certain parameters with specific priors, and ``Marg" represents the results obtained by marginalizing certain parameters with a flat prior. }
\label{otherobser}
\end{center}
\end{table*}

\begin{figure}[htbp]
\includegraphics[width=8cm]{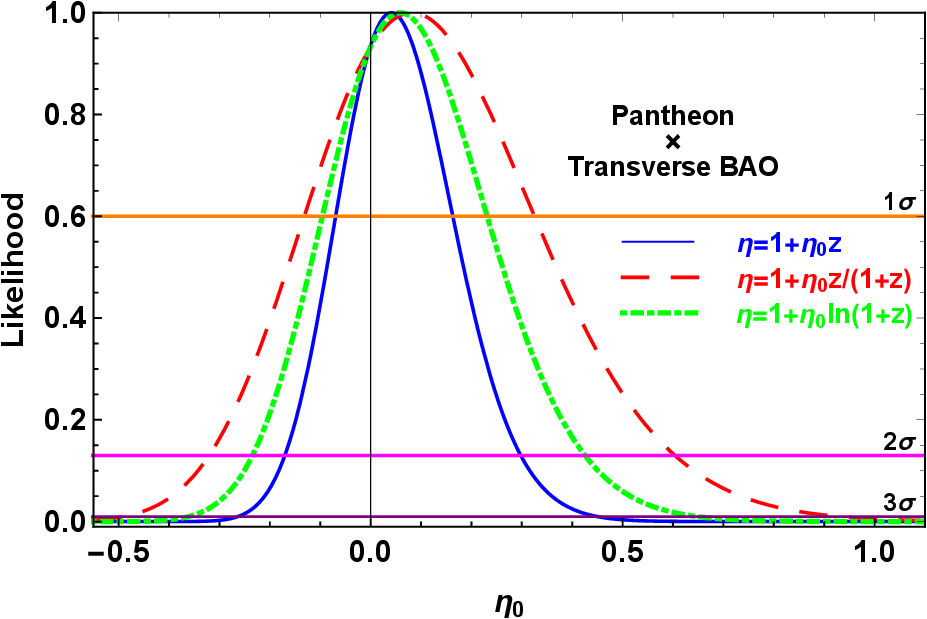}
\caption{\label{figeta2} The likelihood distribution obtained from Pantheon SNIa data and transverse BAO data with a flat prior on $\kappa$. }
\end{figure}

\section{ Results and Analysis}
To show the robustness of nonparametric and parametric methods, we also conduct an unbiased test by generating mock data for $D_{\rm A}$ and $D_{\rm L}$ that strictly adheres to $\eta(z)=D_{\rm L}(z)(1 + z)^{-2}/D_{\rm A}(z)=1$.  The redshifts and uncertainties are taken
from the actual observational Pantheon and transverse BAO data.
The fiducial LDs ($D_{\rm L}^{\rm fid}$) or ADDs ($D_{\rm A}^{\rm fid}$) of the mock data are obtained from a
known flat  $\Lambda$CDM with the most recent Planck results~\cite{Ade2016}. The simulated measurements of $D_{\rm A}$ or $D_{\rm L}$ can be derived
from the random normal distribution through the equation
$D_{\rm L/A}=\mathcal{N}(D_{\rm L/A}^{\rm fid}, \sigma_{D_{\rm L/A}})$. The results are shown in Fig.~\ref{figeta2}, and it can be concluded that CDDR is consistent with the mock data at $1\sigma$ CL from the parametric and nonparametric methods. The possible deviations from the CDDR at any redshifts can be validated using the nonparametric method.
\begin{figure}[htbp]
\includegraphics[width=8cm]{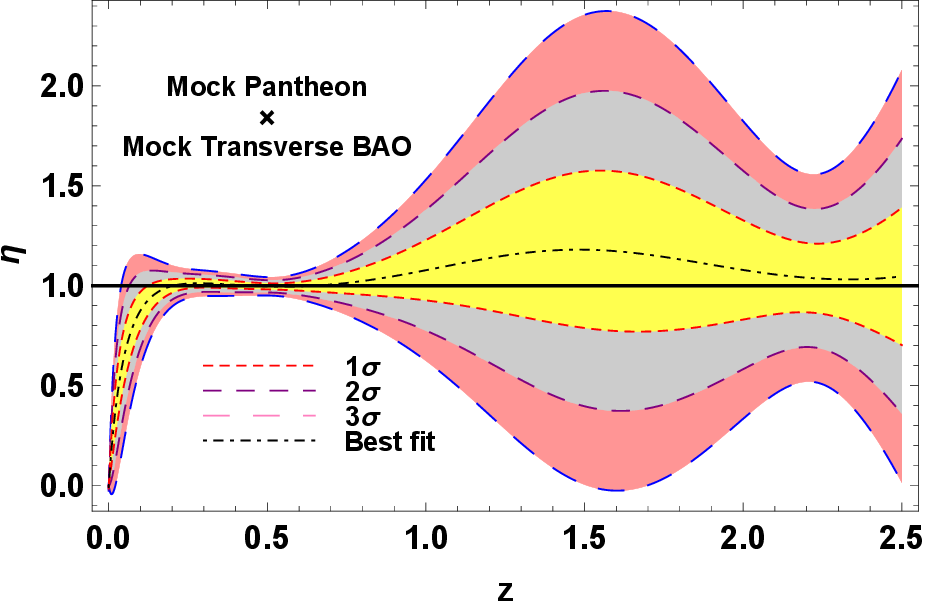}
\includegraphics[width=8cm]{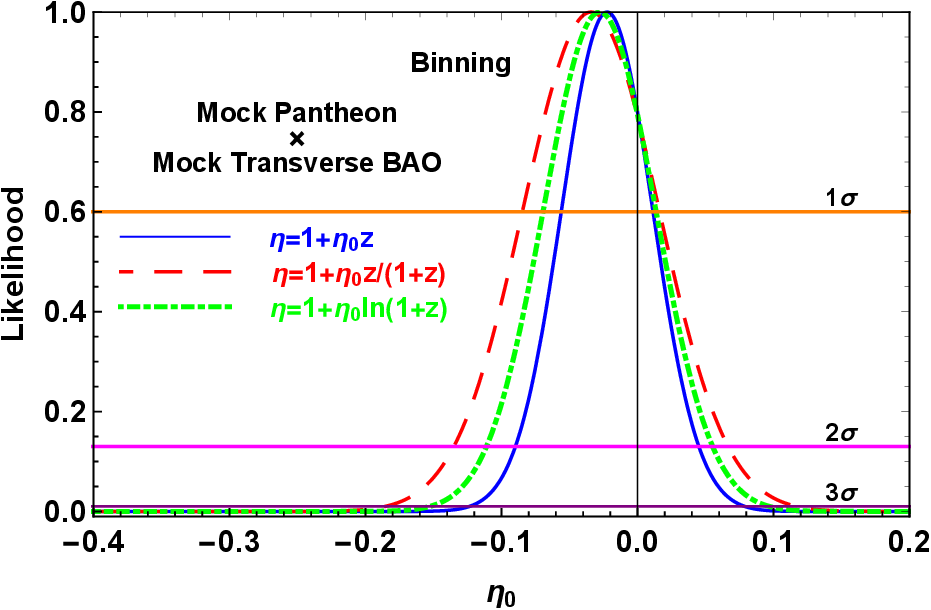}
\caption{\label{figeta2}  The distribution of $\eta(z)$ obtained from the mock data with nonparametric methods (left panel), and the likelihood distribution function obtained from parametric methods (right panel). }
\end{figure}

As shown in Fig.~\ref{figeta1}, for the reconstructed results with the nonparametric methods, within the redshift range $0<z<0.1$, the deviations of the value of function $\eta(z)$ from $1$  may be due to the lack of the
observed transverse BAO data points. CDDR is consistent with the observational data at $2\sigma$ CL with the priors of $M_{\rm B}^{\rm D20}$ and $r_{\rm s}^{\rm C20}$,  at $3\sigma$ CL with the prior of  $M_{\rm B}^{\rm D20}$ and $r_{\rm s}^{\rm V17}$, and at $3\sigma$ CL marginally with the priors of $M_{\rm B}^{\rm B23}$ and $r_{\rm s}^{\rm C20}$, respectively. However, with the priors of $M_{\rm B}^{\rm B23}$ and $r_{\rm s}^{\rm V17}$, the results indicate that there is a strong deviation between the CDDR and the observed data in the redshift region $0.2<z<0.7$ where most of the observed transverse BAO data are located. Thus, the priors of $M_{\rm B}$ and $r_{\rm s}$ will cause significant bias in the CDDR test.

From Fig.~\ref{figeta} and Tab.~\ref{likelihood1}, the results obtained with the three types of parameterizations from the binning method are consistent with that from the nonparametric method. Our results indicate a stronger violation from CDDR than that obtained from Ref.~\cite{Kumar2022}, where the CDDR was fairly consistent with different best fitted value of $M_{\rm B}$ from Pantheon SNIa and Hubble data at $2\sigma$ CL in both flat and a non-flat universe.
Thus, the parametric method provides an effective way to check the CDDR through constraining the parameter $\eta_0$.

Afterwards, we use the Akaike Information Criterion (AIC)~\cite{Akaike1974} and Bayesian Information Criterion~\cite{Schwarz1978} to compare the results obtained from the three parameterizations. The AIC and the
BIC are defined as the following forms, respectively, 
\begin{equation}
\label{AIC}
{\rm AIC}\equiv-2\ln \mathcal{L}_{\rm max}+2p\,,
\end{equation}
\begin{equation}
\label{BIC}
{\rm BIC}\equiv -2\ln \mathcal{L}_{\rm max}+p\ln N\,.
\end{equation}
Here, $\mathcal{L}_{\rm max}$ denotes the value of the maximum likelihood estimate, $p$ represents the number of
free parameters of the model, and $N$ is the total number of data points.
The results are listed in Tab.~\ref{likelihood2}. In all of the cases, the value of $ \chi^2_{\rm min} $  for the parametrization P2 is the smallest one among the three parameterizations, although the parametrization P1 provides the most stringent constraints on the test of CDDR. We can get that all the value of $\Delta{\rm AIC}(\Delta{\rm BIC})< 1.2$, and we cannot determine which parametrization is the most appropriate one, since $\Delta{\rm AIC}(\Delta{\rm BIC})\in [0, 2]$  indicates a weak evidence in favour of
the reference model, which still leaves the question of which model is the most appropriate~\cite{Akaike1974,Schwarz1978}.

Now, we compare the capability of transverse BAO measurements with those obtained from other astronomical observations,  where specific priors of cosmological variables are used. The transverse BAO measurements improve the accuracy of $\eta_0$ about 65\% at $1\sigma$ CL by comparing with the results obtained from the Union2+galaxy cluster observation (the elliptical $\beta$ model)~\cite{Li2011} or  Union2.1+91GMF observation~\cite{Goncalves2015}, and about 40\% at $1\sigma$ CL by comparing with the results obtained from Union2.1+BAO measurements~\cite{Wu2015}, where the CDDR tests were performed with the specific priors of $M_{\rm B}$ or $r_{\rm s}$ from the CMB observations. Our results are about 25\% more stringent  than the  constraints on $\eta_0$ from the SPT-SZ clusters and X-ray measurements from XMM-Newton~\cite{Bora2021}, where the prior of $M_{\rm B}$ and $H_0$ are used. Our results are also comparable with  that obtained from the x-ray GMF of galaxy clusters jointly with SNIa and CMB temperature~\cite{Holanda2017}, where the $M_{\rm B}$ is fixed to derive the LD.

As for the case of testing CDDR with a flat prior of $\kappa=10^{M_{\rm B} \over 5}r_{\rm s}$, the CDDR is
compatible with the observed data at $1\sigma$ CL.  The constraints on $\eta_0$ at $1\sigma$ CL  are much weaker than those
obtained with the specific priors of $M_{\rm B}$ and $r_{\rm s}$. In addition, we marginalize $M_{\rm B}$ and $r_{\rm s}$ separately with the same observational data, and we obtain the same results as those obtained with the method by marginalizing $\kappa$. Therefore, our method of combining two variables into one parameter can reduce the computational complexity arising from integration, and it cannot provide stricter constraints on the parameters. We also perform the CDDR tests with the unanchored LD from Pantheon SNIa while $m_{\rm B}=5\log_{10}[H_0{D_{\rm L}(z)}]-5a_{\rm B}$, where $a_{\rm B}$ is fixed as $a_{\rm B}=0.71273\pm0.00176$ from Ref.~\cite{Riess2016}. We get that the results obtained by marginalizing $a_{\rm B}$ and $r_{\rm s}$ are similar to those obtained by marginalizing $M_{\rm B}$ and $r_{\rm s}$. To explore the capability of testing CDDR with the transverse BAO measurements, it is meaningful to compare our results with the previous constraints on $\eta_0$ from different data sets by marginalizing certain parameters with a flat prior. The transverse BAO measurements improve the accuracy of $\eta_0$ about 50\% at $1\sigma$ CL by comparing with the results obtained from the Union2.1+BAO observations, where the dimensionless Hubble constant $h$ was marginalized with a flat prior~\cite{Wu2015}. The constraints on $\eta_0$ are comparable with that obtained from the latest Pantheon sample and BOSS DR12 BAO measurements in the redshift region $0.31\leq z\leq 0.72$~\cite{Xu2020} or from five BAO measurements from the extended Baryon Oscillation Spectroscopic Survey data release 16 quasar sample
with the Pantheon SNIa samples~\cite{Xu2022} where the $M_{\rm B}$ and $r_{\rm s}$ were marginalized.
Therefore, the ability to test CDDR from the transverse BAO observations is at least comparable to that of previous BAO observations. In addition, it is worth mentioning that in Ref.~\cite{Kumar2022}, the CDDR test was performed through making constraints on $M_{\rm B}$, spatial curvature ($\Omega_{k0}$) and CDDR parameter ($\eta(z)$), resulting different best fitted value $M_{\rm B}$  and spatial curvature ($\Omega_{k0}$). In our analysis, the parametric method used to test CDDR is not only independent of cosmological model, but also independent of the absolute magnitude $M_{\rm B}$ from SNIa observation, sound horizon scale $r_{\rm s}$ of BAO measurements, and Hubble constant $H_0$.

\begin{table}[htp]
\begin{tabular}{cccccccccccc}
\hline\hline
\scriptsize{parmetrization }  & \    \scriptsize{$ \chi^2_{\rm min} $}\ \ & \scriptsize{${\rm AIC}$}  & \   \scriptsize{$ \Delta{\rm AIC}$}  & \   \scriptsize{${\rm BIC}$}   & \   \scriptsize{$\Delta{\rm BIC}$} & \   \scriptsize{parmetrization }   & \   \scriptsize{$\chi^2_{\rm min}$}  & \   \scriptsize{${\rm AIC}$} & \   \scriptsize{$\Delta{\rm AIC}$}  & \   \scriptsize{${\rm BIC}$}  & \   \scriptsize{$\Delta{\rm BIC}$}  \\ \hline
\scriptsize{${\rm P2}^{\rm {A\,C}}\tiny$} & \scriptsize{${12.405}$}& \scriptsize{${18.405}$} & \scriptsize{${0}$} & \scriptsize{${22.179}$} & \scriptsize{${0}$} & \scriptsize{${\rm P2}^{\rm {B\,D}}\tiny$} & \scriptsize{${12.389}$} & \scriptsize{${18.389}$} & \scriptsize{${0}$} & \scriptsize{${22.163}$} & \scriptsize{${0}$}  \\
\scriptsize{${\rm P3}^{\rm {A\,C}}\tiny$} & \scriptsize{${12.448}$}& \scriptsize{${18.448}$} & \scriptsize{${0.043}$} & \scriptsize{${22.222}$} & \scriptsize{${0.043}$} & \scriptsize{${\rm P3}^{\rm {B\,D}}\tiny$} & \scriptsize{${12.886}$} & \scriptsize{${18.886}$} & \scriptsize{${0.497}$} & \scriptsize{${22.660}$} & \scriptsize{${0.497}$} \\
\scriptsize{${\rm P1}^{\rm {A\,C}}\tiny$} & \scriptsize{${12.495}$}& \scriptsize{${18.495}$} & \scriptsize{${0.090}$} & \scriptsize{${22.269}$} & \scriptsize{${0.090}$} & \scriptsize{${\rm P1}^{\rm {B\,D}}\tiny$} & \scriptsize{${13.528}$} & \scriptsize{${19.528}$} & \scriptsize{${1.139}$} & \scriptsize{${23.302}$} & \scriptsize{${1.139}$} \\ \hline
\scriptsize{${\rm P2}^{\rm {B\,C}}\tiny$}  & \scriptsize{${11.084}$}& \scriptsize{${17.084}$} & \scriptsize{${0}$} & \scriptsize{${20.858}$} & \scriptsize{${0}$} & \scriptsize{${\rm P2}\tiny\star$} & \scriptsize{${20.806}$} & \scriptsize{${22.806}$} & \scriptsize{${0}$} & \scriptsize{${24.064}$} & \scriptsize{${0}$} \\
\scriptsize{${\rm P3}^{\rm {B\,C}}\tiny$}  & \scriptsize{${11.128}$}& \scriptsize{${17.128}$} & \scriptsize{${0.044}$} & \scriptsize{${20.902}$} & \scriptsize{${0.044}$} & \scriptsize{${\rm P3}\tiny\star$} & \scriptsize{${20.807}$} & \scriptsize{${22.807}$} & \scriptsize{${0.001}$} & \scriptsize{${24.065}$} & \scriptsize{${0.001}$} \\
\scriptsize{${\rm P1}^{\rm {B\,C}}\tiny$}  & \scriptsize{${11.194}$}& \scriptsize{${17.194}$} & \scriptsize{${0.110}$} & \scriptsize{${20.968}$} & \scriptsize{${0.110}$} & \scriptsize{${\rm P1}\tiny\star$} & \scriptsize{${20.812}$} & \scriptsize{${22.812}$} & \scriptsize{${0.006}$} & \scriptsize{${24.070}$} & \scriptsize{${0.006}$} \\ \hline
\scriptsize{${\rm P2}^{\rm {A\,D}}\tiny$} & \scriptsize{${12.370}$}& \scriptsize{${18.370}$} & \scriptsize{${0}$} & \scriptsize{${22.144}$} & \scriptsize{${0}$} & \scriptsize{${\rm P2}^{\rm {M}}\tiny$} & \scriptsize{${5.860}$} & \scriptsize{${7.860}$} & \scriptsize{${0}$} & \scriptsize{${9.118}$} & \scriptsize{${0}$} \\
\scriptsize{${\rm P3}^{\rm {A\,D}}\tiny$} & \scriptsize{${12.470}$}& \scriptsize{${18.470}$} & \scriptsize{${0.100}$} & \scriptsize{${22.244}$} & \scriptsize{${0.100}$} & \scriptsize{${\rm P3}^{\rm {M}}\tiny$} & \scriptsize{${5.869}$} & \scriptsize{${7.869}$} & \scriptsize{${0.009}$} & \scriptsize{${9.127}$} & \scriptsize{${0.009}$} \\
\scriptsize{${\rm P1}^{\rm {A\,D}}\tiny$} & \scriptsize{${12.611}$}& \scriptsize{${18.611}$} & \scriptsize{${0.241}$} & \scriptsize{${22.385}$} & \scriptsize{${0.241}$} & \scriptsize{${\rm P1}^{\rm {M}}\tiny$} & \scriptsize{${5.878}$} & \scriptsize{${7.878}$} & \scriptsize{${0.018}$} & \scriptsize{${9.136}$} & \scriptsize{${0.018}$} \\
\hline\hline
\end{tabular}
\caption{The results obtained from the AIC and BIC information criterion. The superscripts A, B, C, and D represent the cases obtained from $M_{\rm B}^{\rm D20}$, $M_{\rm B}^{\rm B23}$, $r_{\rm s}^{\rm C20}$, and $r_{\rm s}^{\rm V17}$, respectively. The superscript $\star$ denotes the results obtained from the flat prior for $\kappa$. The superscript M denotes the results obtained from the mock SNIa data and mock BAO data.}
\label{likelihood2}
\end{table}

\section{conclusion}
CDDR plays an important role in astronomic
observations and modern cosmology, and any deviation from CDDR  may imply a signal of exotic physics.
SNIa and BAO measurements can be considered effective observational data for testing CDDR, as they can provide precise LD and ADD data. Due to the uncertainty in determining the absolute magnitude $M_{\rm B}$ from different astronomical observations, as well as the uncertainty in the sound horizon scale $r_{\rm s}$ in the fitting problem of BAO observation, it is necessary to investigate the impact of the quantities $M_{\rm B}$ and $r_{\rm s}$ from SNIa and BAO observations on the CDDR test, and use new BAO measurements and more reliable methods to check the validity of the CDDR.

In this work, we first, use a nonparametric method to test CDDR by comparing the LD obtained from the Pantheon SNIa compilation with ADD from transverse BAO measurements. The function $\eta(z)=D_{\rm L}(z)(1 + z)^{-2}/D_{\rm A}(z)$ is adopted to probe the possible deviations from CDDR at any redshift.
We use the two prior values of the absolute magnitude $M_{\rm B}$ and sound horizon scale $r_{\rm s}$, respectively, to obtain the LD of the SNIa observation and ADD of the transverse BAO measurements. The smoothing functions of LD $D_{\rm L}(z)$ and ADD $D_{\rm A}(z)$ are reconstructed with the Gaussian process. The results show that the different priors of $M_{\rm B}$ and $r_{\rm s}$ will cause significant bias in testing CDDR.

Then, we employ three parameterizations to describe the
function $\eta(z)$ that evolves with the redshift $z$, namely P1: $\eta(z)=1+\eta_0z$, P2: $\eta(z)=1+\eta_0z/(1+z)$, and P3: $\eta(z)=1+\eta_0\ln(1+z)$, and
test CDDR by constraining the parameter $\eta_0$ with the priors of $M_{\rm B}$ and $r_{\rm s}$. Our results show
that P1 offers the best rigorous constraints on the
parameter $\eta_0$ among the three parameterizations. Compared with the results from the
nonparametric method, the same results are obtained from
the nonparametric method for the observational data and
the priors of $M_{\rm B}$ and $r_{\rm s}$. Therefore, the
parametric method offers an effective way to test
CDDR by constraining the parameter $\eta_0$. The
results also imply that in the test of CDDR,
some bias might be caused by the priors of $M_{\rm B}$ and $r_{\rm s}$ when the exact values of observation $M_{\rm B}$ and $r_{\rm s}$ are not determined.

To avoid the bias caused by the priors of $M_{\rm B}$ and $r_{\rm s}$ in the CDDR test, we consider fiducial values of $M_{\rm B}$ and $r_{\rm s}$ as nuisance parameters to determine the LD $D_{\rm L}$ and ADD $D_{\rm A}$ from the astronomic observations, and then marginalize their influences with a
flat prior of new variable $\kappa\equiv 10^{M_{\rm B} \over 5}r_{\rm s}$ in the analysis.  We match the SNIa data with the transverse BAO measurements at the same redshift with the binning method. Our results show that CDDR
is consistent with the current observational data at $1\sigma$ CL,
and the ability of observational data with this method to constrain the parameter $\eta_0$ is weaker
than that obtained from the specific values of $M_{\rm B}$ and $r_{\rm s}$. The reason is that there is one more degree of freedom in our analysis compared to the previous analysis. Compared with previous results from other astronomic observations, the capability of the transverse BAO observation to test the CDDR is at least comparable to previous observations, whether using the method  independent of $M_{\rm B}$ and $r_{\rm s}$ or the method dependent of $M_{\rm B}$ and $r_{\rm s}$. It is meaningful to note that this method to test CDDR is not only independent of cosmological models, but also independent of the prior values of absolute magnitude $M_{\rm B}$ and sound horizon scale $r_{\rm s}$.  In addition, with the developments of the BAO measurements, the neutral hydrogen intensity mapping
technique can be used to measure the BAO signals more
efficiently, and 200 observational data in the redshift region $0<z<2.5$ will
be realized in the coming decades~\cite{Zhang2020}. As the
quality and quantity of measurements for SNIa
and BAO measurements increase, the parametric method in
our analysis will be a powerful way to test CDDR  independent of both the cosmological model and
the values of $M_{\rm B}$ and $r_{\rm s}$. Thus, the transverse BAO measurements can be used as a powerful tool to test CDDR.

\begin{acknowledgments}
We very much appreciate helpful comments and suggestions from anonymous referees, and helpful discussion from Hongwei Yu and Puxun Wu. This work was supported by the National Natural Science Foundation of China under Grants No. 12375045,
No. 12305056, No. 12105097, and No. 12205093, the
Hunan Provincial Natural Science Foundation of China
under Grants No. 12JJA001 and No. 2020JJ4284, the Natural Science Research Project of Education
Department of Anhui Province No. 2022AH051634, and the
Science Research Fund of Hunan Provincial Education
Department No. 21A0297.

\end{acknowledgments}

\end{document}